# Interatomic potentials for the vibrational properties of III-V semiconductor nanostructures

Peng Han and Gabriel Bester*

Max-Planck-Institut fuer Festkoerperforschung, Heisenbergstr. 1, D-70569 Stuttgart, Germany

We derive interatomic potentials for zinc blende InAs, InP, GaAs and GaP semiconductors with possible applications in the realm of nanostructures. The potentials include bond stretching interaction between the nearest and next-nearest neighbors, a three body term and a long-range Coulomb interaction. The optimized potential parameters are obtained by (i) fitting to bulk phonon dispersions and elastic properties and (ii) constraining the parameter space to deliver well behaved potentials for the structural relaxation and vibrational properties of nanostructure clusters. The targets are thereby calculated by density functional theory for clusters of up to 633 atoms. We illustrate the new capability by the calculation Kleinman and Grüneisen parameters and of the vibrational properties of nanostructures with 3 to 5.5 nm diameter.

## 1   Introduction

The study of confinement effects on the electronic and optical properties of nanostructure represents one of the most vibrant fields of condensed matter physics. While the confinement effects on electronic and optical properties of these structures have been intensively investigated and are fairly well understood, much less is known about their vibrational properties. The phonons from the bulk with their well studied dispersions collapse into discrete states, or vibrons[1] in the case of 3D confinement given in quantum dots (QDs). These excitations allow for carriers to relax down the discrete ladder of electronic or excitonic states. For the larger self-assembled quantum dots the electronic and excitonic states are believed to enter a strong coupling regime with the phonons to form polarons[2], [3]. They are proposed as explanation for the lack of phonon-bottleneck[4]. In colloidal nanocrystals the reasons for the lack of phonon-bottleneck is still under debate[5], [6] but is likely to involve vibrons. Vibrational properties are also most relevant to the loss of quantum coherence; a process which limits the application of quantum dots in the field of quantum optics and information science.

The phonon density of states (DOS) and dispersion of bulk semiconductors can be calculated[7], [8] with great accuracy via ab initio density-functional perturbation theory (DFPT)[9], [7]. Ab initio study on the vibrational properties of semiconductor nanostructures such as fullerenes[10], nanowires[11], [12], nanotubes[13], and nanoclusters with small sizes[14], [15] have been reported. Although an excellent description of vibrational properties can be achieve by density-functional theory (DFT) based calculations, the state-of-the-art DFT method is limited in the praxis to systems with translational periodicity (3-dimensional, 2-dimensional, or 1-dimensional) or very small 0-dimensional QDs, due to the high computational demand.

In order to calculate vibrons for high quality nanostructures, one must be able to address diameters ranging from several to tens of nanometers. Indeed, smaller structures tend to exhibit poor optical characteristics and show large variations within one sample. Furthermore, the recent trend in colloidal chemistry is to produce core-shell structures, which are rather large. Although these structures have several thousand atoms, the surface or interface atoms represent a significant fraction of the total atom number. In a colloidal structure with 1000 atoms one quarter of the atoms are directly on the surface. Hence, the use of continuum dielectric models based on bulk phonons are expected to represent only a poor approximation. This situation calls for a reliable and cheap interatomic potential for the vibrational properties.

## 2 The necessity to be able to relax the structure to avoid imaginary frequencies

One of our prerequisite for the derivation of an empirical potential for the calculation of vibrons is the ability to use the potential to relax the structure. This is motivated by the fact that an unrelaxed structure will lead to the appearance of unphysical imaginary frequencies. Based on the harmonic approximation of lattice dynamics, the phonon frequencies $\omega$ and the vibrational eigenvectors $a_{k\alpha}(\mathbf{q})$ with phonon wave vector $\mathbf{q}$ are obtained by solving the eigenvalue equation

$$\sum_{k'\beta} D^{k\alpha}_{k'\beta}(\mathbf{q}) a_{k'\beta}(\mathbf{q}) = \omega^2 a_{k\alpha}(\mathbf{q})$$

(1)

where the dynamical matrix $D$ is given by:

$$D^{k\alpha}_{k'\beta}(\mathbf{q}) = \sum_l \frac{1}{\sqrt{m_k m_{k'}}} \frac{\partial^2 V}{\partial u_{0k\alpha} \partial u_{lk'\beta}} e^{i\mathbf{q}\cdot\mathbf{r}_l},$$

(2)

where, $l$ and $k$ are the unit cell and atomic index, respectively, and $\alpha$ is the Cartesian direction, $m_k$ is the mass of atom $k$, $V$ is the potential energy of the crystal, $\frac{\partial^2 V}{\partial u_{0k\alpha} \partial u_{lk'\beta}}$ is the force constant matrix element. The dynamical matrix is Hermitian and hence the eigenvalues real and the frequencies either purely real or purely imaginary.

For structures with well relaxed geometry, all the atoms occupy their equilibrium positions with the lowest potential energy. In this case, the values of the first order derivative of the potential energy (forces, $-\frac{\partial V}{\partial u_{lk\alpha}}$) are equal to zero, while those of the second order derivatives (force constant matrix elements) are positive and hence all the eigenvalues of Eq. ((1)) (square of vibrational frequencies, $\omega^2$) are positive. This leads to well defined (real) vibrational frequencies. In the case of an unrelaxed structure the atoms are displaced from their equilibrium positions in the potential landscape and the first and second order derivates may have any sign. Hence the eigenvalues of the dynamical matrix may be negative, leading to imaginary frequencies.

The need to relax the structure becomes coercive when one wants to address issues such as the effect of interfaces or surfaces on vibrations. One common strategy in such situations is to use existing potentials (see below) known to lead to good phonon dispersions in the bulk and to relax the structure via other means (DFT or valence force field (VFF))[16], [17], [18]. However, this leads to a somewhat inconsistent treatment, where the relaxed input structure does not represent the lowest strain energy structure within the phonon-model potential. Again, imaginary frequencies ensue from such an approach.

## 3 The need for long range interactions

The most striking deficiencies of empirical models without long range interaction is the failure to describe the splitting of the longitudinal and transverse optical (LO-TO) phonons at the zone center and the failure to describe the flattening of the transverse acoustic (TA) phonon at the boundary of the Brillouin zone. In Fig. 1 we show, as illustration for the effect of a missing long-range term, results of a valence force field calculation (black solid lines) and DFPT (red circles). The above mentioned deficiencies are noticeable, as well as a blue shift of the optical branches. Note that the potential parameters used in Fig. 1 reproduce the elastic properties rather well[19]. Our conclusion is that existing potentials reproduce either the elastic properties or the phononic properties well, but never both, as suggested earlier. Deficiencies, such as shown in Fig. 1 can be partially (e.g., the blue shift of the optical branches, not the LO-TO splitting) remedied by an adjustment of parameters within the same model, but at the expense of the quality of the elastic properties.

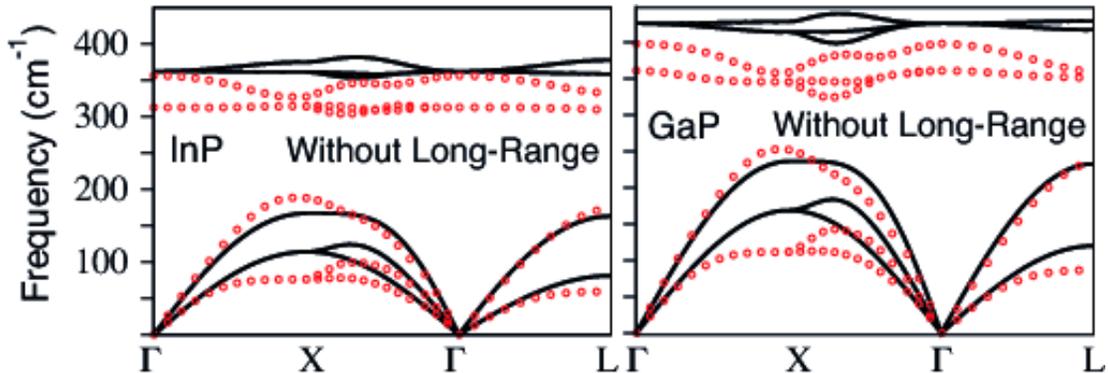

**Figure 1: (Color Online) Black solid lines: Phonon dispersion calculated using a Tersoff potential (without long-range Coulomb interaction) for InP and GaP and (circles) via DFPT.**

## 4  Existing Empirical Models for Phonons

The empirical description of bulk phonons in semiconductors has a long history.The most famous models can be cast into the categories of the shell models[20], the bond charge models (BCM)[21], [22] and the VFF [23], [24].

While the shell models are historically important, only few recent applications have been reported, due to the severe underlying approximation of a spherical distribution of covalently bonded electrons. Furthermore, the parameters determined from these types of models lack an obvious physical interpretation.

The BCMs can be parametrized to lead to bulk phonon dispersions with an accuracy comparable to DFPT and can be seen as the most successful empirical models for bulk phonons. However, the atomic forces, necessary to obtain a structurally relaxed and stable structure are not directly available from BCMs. In BCMs, the force constants rather than the potential are the central quantities. This is unproblematic for bulk calculations but leads to imaginary frequencies in nanostructures, as mentioned above. A way to obtain forces is to assume a certain functional form for the ion-ion potential[25], such as a Born-Mayer potential[25], and extract the parameters for this potential from the BCM parameters. This is a viable approach, but leads to the awkward situation to have different models for the structural relaxation and for the calculation of phonons.

In the VFF models, the potential energy of the valence bonds is expanded phenomenologically in terms of valence bond parameters, which correspond to bond stretching and bond bending interactions. A simplified version of VFF model was derived by Keating[23] leading to a simple but accurate description of the elastic properties of covalently bonded semiconductors. One of the appeal of the method is the direct link between the (measurable) elastic constants and the model parameters. However, the original model [23] lacks a long-range interaction and therefore fails to describe the splitting of the LO-TO phonon branches at the zone center and fails to describe the flattening of the TA phonon branch at the boundary of the Brillouin zone. The flattening of the TA branch can be improved by introducing additional parameters[26] but polarization effects require a long-range interaction. An additional long range Coulomb potential was soon introduced into the VFF model[24]. However, within this model, the parameters which fit the elastic properties do not reproduce phonons well[23] and vice versa[27], [28].

Another type of very popular potentials are based on the parametrization given by Stillinger and Weber[29]. These potentials were derived in order to allow for an approximate treatment of bond breaking as typically encountered in molecular dynamic simulations, originally for liquid Silicon. However, for tetrahedrally coordinated materials these type of potentials do not lead to an improvement over the Keating model, as already stated in the original work of Stillinger and Weber[29].

The state-of-the-art preferred choice for empirical atomistic modeling of structural properties of semiconductor surfaces and interfaces are bond order potentials (BOP)[30], [31]. One of them is the Abel-Tersoff potential (ATP)[30], [32], [33], [34], [19], [35], [36] originally proposed[30] to describe accurately the properties of non-tetrahedral forms of silicon. It has been successively improved several times and extended to the group IV[37], [38], [39] and III-V semiconductors[32], [33], [34], [19], [35], [36]. The original BOPs have also branched off, while keeping the same name, into a different kind of BOPs based on tight-binding and the recursion method [31]. The BOPs are characterized by a many body term with a functional dependence on the bond order of the local environment. This provides a simple environmental dependence and some short-range non-locality to the potential. It allows the treatment of non-tetrahedral sites at the surface of semiconductors. However, the long range Coulomb interaction, which is critical for a proper phonon description, is difficult to incorporate into the ATP framework.

## 5 New interatomic potential

The potential we propose in the present work consists of short-range two-body terms $V_2(i,j)$, a three-body term $V_3(i,j,k)$ and a long-range Coulomb term $V_C(i,j)$. The two-body terms, which describe the short-range bond stretching interactions between the first nearest neighbor(1NN) and second nearest neighbor (2NN) atoms, have the following form:

$$V_2(i,j) = \begin{cases} A(\dfrac{B}{r_{ij}^4}-1)\exp(\dfrac{1}{r_{ij}-b}), & r_{ij} < b \\ 0, & r_{ij} \geq b \end{cases} \tag{3}$$

where, $r_{ij}$ denotes the distance between atom $i$ and $j$, $b$ represents the cutoff distance of the interaction. We chose this functional form because it allows for bond-breaking at the surface and also for a practical reason: it is equivalent to the form suggested by Stillinger and Weber[29] (SW) and is available in most empirical potential codes (e.g. IMD [40], LAMMPS [41], GULP [42]) as one of the possible choices of potential. Important to us is the harmonic behavior for small atomic displacement which is, of course, given by the SW form. We do not claim a meaningful non-harmonic behavior of our potential, which we consider to be beyond the capability of empirical potentials. The parameters $A$ and $B$ affect the bond-strength and bond-length, respectively.

The three-body term corresponds to the contribution of an angle distortion with respect to the ideal tetrahedral angle and is given by:

$$V_3(i,j,k) = h(r_{ij},r_{ik}) + h(r_{ji},r_{jk}) + h(r_{ki},r_{kj}), \tag{4}$$

where

$$h(r_{ij},r_{ik}) = \lambda \exp(\dfrac{\eta}{r_{ij}-b} + \dfrac{\eta}{r_{ik}-b})(\cos\theta_{jik} + \dfrac{1}{3})^2. \tag{5}$$

In Eq. ((5)), $r_{ik}$ represents the distance between atom $i$ and $k$, $\theta_{jik}$ denotes the angle subtended by $r_{ij}$ and $r_{ik}$ with the vertex at $i$, $\lambda$ and $\eta$ are potential parameters. As shown in Eq. ((5)), the three-body interaction will be zero at the ideal tetrahedral angle $\theta_t = 109.49°$ and is positive otherwise. We use an additional long-range Coulomb interaction between atoms (in atomic units):

$$V_C(i,j) = \frac{Z_i^* Z_j^*}{|r_{ij}|}, \tag{6}$$

where $Z_i^*$ and $Z_j^*$ denote the effective charges of the atoms $i$ and $j$, respectively. Considering both the short- and long-range interactions, the total force field potential energy is

$$V = \sum_{i<j}^{j \in 1NN} V_2(i,j) + \sum_{i<k}^{k \in 2NN} V_2(i,k) + \sum_{i<j<k} V_3(i,j,k) + \sum_{i<j} V_C(i,j). \tag{7}$$

The forces, the force constants, and the dynamical matrix elements can be calculated with Eq. ((7)). Thereafter, the phonon modes are obtained by diagonalizing the dynamical matrix.

The physical reason for the introduction of a 2NN interaction can be understood from Fig. 2. In Fig. 2, the total and the Coulomb interactions between the 1NN and 2NN atoms are shown for (a) InAs, (b) InP, (c) GaAs, and (d) GaP. As shown in these figures, the first order derivative of the total potential energy between the 2NN atoms is nearly flat at the equilibrium position of the 2NN. This is the consequence of a near cancellation, or screening, of the Coulomb repulsion via the 2NN attractive interaction. With this effective screening we can obtain a potential that includes a long-range Coulomb term but that properly binds atoms at the surface.

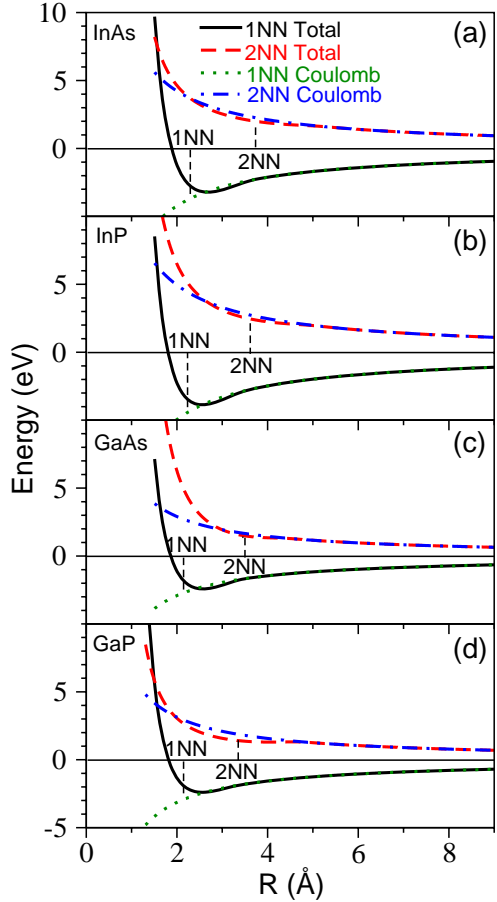

**Figure 2: (Color online) The plots of the total interactions between 1NN atoms (solid lines), 2NN atoms (dashed lines), Coulomb interactions between 1NN atoms (dotted lines), and 2NN atoms (dashed and dotted lines) for (a) InAs (b) InP, (c) GaAs, and (d) GaP.**

## 6  Fitting method

The parameters for this potential are obtained by fitting them to reproduce the bulk-phonon dispersion obtained from ab initio calculations and the experimental elastic constants $C_{11}$, $C_{12}$ and $C_{44}$. The target bulk-phonon dispersion used in the fitting procedure are calculated via DFPT, as implemented in ABINIT[43]. In the calculation, we employ the local density approximation (LDA), Trouiller-Martin norm-conserving pseudopotentials with plane-wave expansion up to a 30 Ry cutoff. The phonon frequencies are converged up to ± 0.1 cm$^{-1}$ with a cutoff of 30 Ry and a Monkhorst-Pack mesh of $8\times8\times8$ k-points. The DFPT calculated phonon dispersion (red circles) along the symmetry line $\Gamma \to X \to K \to \Gamma \to L$, together with the corresponding phonon DOS (red thin curves) of InAs, InP, GaAs, and GaP are plotted in Fig. 3 (a)-(h). The experimental and LDA equilibrium lattice constants, along with the experimental and the calculated optical phonon frequencies at the $\Gamma$ point are given in Table. 1. As shown in this table, an excellent agreement for the LO and TO frequencies can be obtained when the LDA lattice constant is used in the DFPT calculations. The agreement is significantly worse at the experimental lattice constant, so that we use LDA lattice constants in all ab initio calculations.

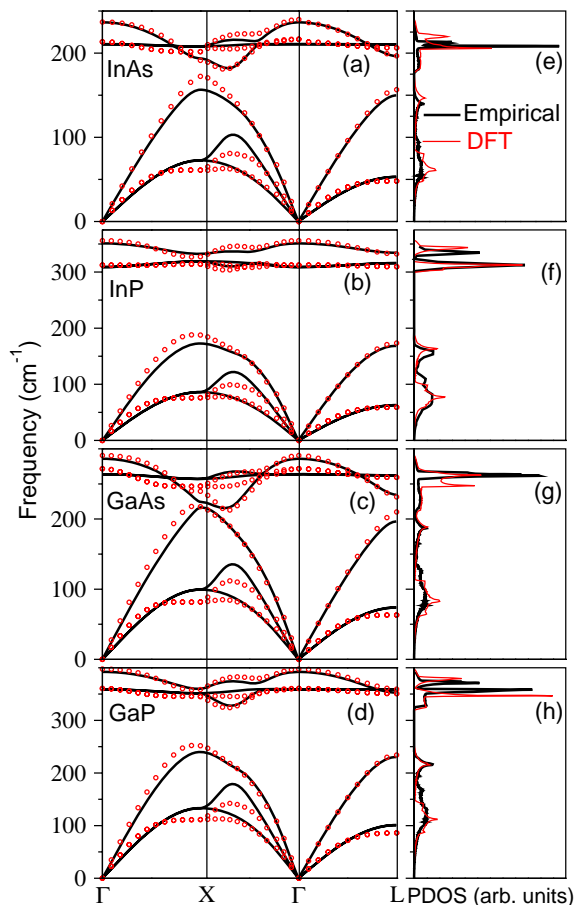

**Figure 3: (Color online) Phonon dispersion curves calculated with the modified SW potentials (solid lines) and with ABINIT (circles) for (a) InAs, (b) InP, (c) GaAs, and (d) GaP. Phonon DOS calculated with the modified SW potentials (thick lines) and with ABINIT (thin lines) for (e) InAs, (f) InP, (g) GaAs, and (h) GaP.**

**Table 1: Experimental lattice constant $a^{\exp}$, and LDA lattice constant $a^{LDA}$ in Å. Experimental and LDA calculated LO and TO phonon frequencies (in cm$^{-1}$) at the $\Gamma$ point calculated at the experimental lattice constant $\omega_{LO}^{LDA}(a=a^{\exp})$ and at the LDA lattice constant $\omega_{LO}^{LDA}(a=a^{LDA})$.**

|  | InAs | InP | GaAs | GaP |
|---|---|---|---|---|
| $a^{exp}$ | 6.06 | 5.86 | 5.65 | 5.45 |
| $a^{LDA}$ | 5.93 | 5.71 | 5.56 | 5.38 |
| $\omega_{LO}^{exp}$ | 240 | 350 | 293 | 402 |
| $\omega_{LO}^{LDA}(a=a^{LDA})$ | 236 | 354 | 290 | 399 |
| $\omega_{LO}^{LDA}(a=a^{exp})$ | 209 | 309 | 269 | 378 |
| $\omega_{TO}^{exp}$ | 218 | 308 | 271 | 366 |
| $\omega_{TO}^{LDA}(a=a^{LDA})$ | 213 | 310 | 271 | 361 |
| $\omega_{TO}^{LDA}(a=a^{exp})$ | 187 | 267 | 250 | 341 |

The fitting procedure uses a Newton-Raphson functional minimization, as is implemented in GULP[42]. The phonon frequencies of 41 k-points in the BZ along $\Gamma \to X \to K \to \Gamma \to L$, and the elastic constants of $C_{11}$, $C_{12}$ and $C_{44}$ are used as weighted targets properties in the fit. The inclusion of elastic constants in the fit is crucial to guarantee a reasonable behavior during structural relaxation of the nanostructure. To obtain a good balance between elastic properties and phonon DOS, the relative weights of the elastic constants are set as 25 times to those of the phonon frequencies. In order to accelerate the convergence of the fitting procedure we used the potential parameters given in Ref.[44] as initial guess.

## 7 Results for bulk and comparison with DFT

The obtained potential parameters are given in Table. 2. For the charges in the Coulomb term, the relation $Z_{anion}^* = -Z_{cation}^*$ is enforced, but the sign of $Z^*$ is undefined. However, we find that the obtained charges follow the electronegativity of the elements: $\chi_P > \chi_{As} > \chi_{Ga} > \chi_{In}$ [45]. Indeed, from the largest to the smallest electron transfer (or ionicity) we obtain as result of the fit (see Tab. 2), InP (0.83), InAs (0.77), GaP (0.66) and GaAs (0.64) in agreement with the electronegativity scale. We attribute positive charges to the cation (In,Ga) and negative charges to the anions (As,P).

**Table 2: Optimized parameters for the empirical interatomic potentials**.

|  | InAs | InP | GaAs | GaP |
|---|---|---|---|---|
| $A_{1NN}$ (eV) | 4.4074 | 5.7630 | 4.6264 | 2.9992 |
| $B_{1NN}$ (Å$^4$) | 48.3326 | 43.1490 | 35.7999 | 47.8668 |
| $b_{1NN}$ (Å) | 4.1074 | 3.9683 | 3.8416 | 3.7156 |
| $A_{2NN}$ (eV) | 1.0195 | 1.4902 | 1.7595 | 1.5970 |
| $B_{2NN}$ (Å$^4$) | 28.0499 | 50.0499 | 80.0499 | 14.3639 |
| $b_{2NN}$ (Å) | 5.6400 | 5.5440 | 4.8830 | 5.1996 |
| $\lambda$ (eV) | 27.9402 | 23.0402 | 30.0402 | 30.6346 |
| $\eta$ (Å) | 2.8697 | 2.7455 | 2.5610 | 2.3371 |
| $|Z^*|$ | 0.7663 | 0.8289 | 0.6355 | 0.6603 |

The bulk phonon dispersion and phonon DOS calculated with the new potentials and via DFPT for InAs, InP, GaAs, and GaP are given in Fig. 3 (a)-(h). As shown in Fig. 3 (a)-(d), the phonon branches agree well along the $\Gamma \to$ L direction. In the $\Gamma \to$ X and $\Gamma \to$ K $\to$ X directions, the empirical results also exhibit a good agreement with the DFPT results except for the TA branches around the X point. The agreement of the TA branch around the X point can be improved using a bond charge model. In the BCM, bond charges are used to describe the interaction of the valence electrons in the bonding region with the ionic cores. With this ingredient, the behavior of the TA branches at the edge of the BZ can be significantly improved [22]. However, we renounce to introduce such a term, which would increase the number of parameters for a marginal improvement. Fig. 3 (e)-(h) shows the phonon DOSs calculated by using our new potential and DFPT, which exhibit a good agreement throughout the frequency range.

The comparison between the experimental and calculated lattice and elastic constants for zinc blende InAs, InP, GaAs, and GaP are listed in Table. 3. In this table, the label $t$ represents the experimental data used as target in the fitting procedure, while $c$ denotes the calculated results by using the empirical potential. As shown in the first column of Table. 3, the relaxed lattice constants calculated with the new potentials agree well with the experimental values. The agreement between the calculated elastic constants of the relaxed structure and the experimental values is also reasonably good, with some errors in $C_{44}$.

**Table 3: Comparison of lattice constants and elastic constants between the target and calculated result.**

|            | $a_0$ (Å) | $C_{11}$ (dyn/cm$^2$) | $C_{12}$ (dyn/cm$^2$) | $C_{44}$ (dyn/cm$^2$) |
|------------|-----------|-----------------------|-----------------------|-----------------------|
| InAs$^t$   | 6.058     | 8.34×10$^{11}$        | 4.54×10$^{11}$        | 3.95×10$^{11}$        |
| InAs$^c$   | 6.053     | 7.54×10$^{11}$        | 4.67×10$^{11}$        | 2.16×10$^{11}$        |
| InP$^t$    | 5.868     | 10.1×10$^{11}$        | 5.61×10$^{11}$        | 4.56×10$^{11}$        |
| InP$^c$    | 5.871     | 9.87×10$^{11}$        | 6.68×10$^{11}$        | 2.59×10$^{11}$        |
| GaAs$^t$   | 5.650     | 11.9×10$^{11}$        | 5.34×10$^{11}$        | 5.96×10$^{11}$        |
| GaAs$^c$   | 5.661     | 10.7×10$^{11}$        | 6.02×10$^{11}$        | 3.36×10$^{11}$        |
| GaP$^t$    | 5.450     | 14.1×10$^{11}$        | 6.20×10$^{11}$        | 7.03×10$^{11}$        |
| GaP$^c$    | 5.441     | 12.9×10$^{11}$        | 6.39×10$^{11}$        | 4.43×10$^{11}$        |

For the diamond and zinc blende structures, the bond length between the two atoms inside the primitive cell under a [111] strain is undetermined and an internal strain can develop[46]. A parameter $\zeta$ quantifying this internal strain was introduced by Kleinman[46] describing the relative ease of bond bending versus the bond stretching. Minimizing bond bending leads to $\zeta = 0$, minimizing bond stretching leads to $\zeta = 1$. Later, Harrison linked the Kleinman parameter in an approximated way to the elastic constants $C_{11}$ and $C_{12}$:

$$\zeta = \frac{C_{11} + 8C_{12}}{7C_{11} + 2C_{12}}.$$

(8)

We give the experimental value and our empirical potential value, following consistently Eq. ((8)) in both cases, in Table 4. As shown in this table, our potentials give slightly higher values of $\zeta$ than those from experimental elastic constants**Error! Reference source not found.** for all the materials. This indicates that we somewhat overestimate the bond stretching term compared to the bond bending term [46].

**Table 4: Kleinman parameters $\zeta$, Cohesive energy $E_{coh}$ and Grüneisen parameters $\gamma$, calculated from our potential (calc.) and taken from experiments (exp.). The Kleinman parameters are all calculated via Eq. ((8)) using theoretical (calc.) or experimental (exp.) elastic constants and therefore represent an approximation.**

|  | InAs | InP | GaAs | GaP |
|---|---|---|---|---|
| $\zeta$ (calc.) | 0.723 | 0.768 | 0.677 | 0.621 |
| $\zeta$ **Error! Reference source not found.** (exp.) | 0.666 | 0.672 | 0.581 | 0.574 |
| $E_{coh}$ (calc.) | -3.60 | -4.33 | -2.60 | -3.40 |
| $E_{coh} Harrison80$ (exp.) | -3.10 | -3.48 | -3.25 | -3.56 |
| $\gamma^{\Gamma}_{TO}$ (calc.) | 0.99 | 1.02 | 0.99 | 1.05 |
| $\gamma^{\Gamma}_{TO} aoki84$ (exp.) | 1.21 | 1.44 | 1.39 | 1.09 |
| $\gamma^{\Gamma}_{LO}$ (calc.) | 0.89 | 0.91 | 0.92 | 0.96 |
| $\gamma^{\Gamma}_{LO} aoki84$ (exp.) | 1.06 | 1.24 | 1.23 | 0.95 |
| $\gamma^{L}_{TA}$ (calc.) | -0.53 | -0.53 | 1.25 | 0.11 |
| $\gamma^{L}_{TA} Yu10$ (exp.) |  | -2.00 | -1.70 | -0.81 |
| $\gamma^{L}_{TO}$ (calc.) | 0.99 | 1.02 | 1.08 | 0.98 |
| $\gamma^{L}_{TO} Yu10$ (exp.) |  | 1.40 | 1.50 | 1.50 |

The cohesive energies $E_{coh}$ obtained experimentally and with our empirical potentials are listed in Table 4 showing qualitative agreement. We did not include the cohesive energies in the fit to avoid a further bias towards better elastic properties and worse phononic properties.
The mode Grüneisen parameters

$$\gamma_m = -\frac{d\ln\omega_m}{d\ln V} \qquad (9)$$

of bulk InAs, InP, GaAs, and GaP, which are calculated from the dependence of the $m$ phonon mode frequency $\omega_m$ on the change of volume $V$, are listed in Table. 4 with the corresponding values obtained from the literatures[47]. Although the deviation of the mode Grüneisen parameter from zero is a consequence of the non-harmonicity of the potential we obtain reasonable values for the optical modes. However, our simple model potential fails for the zone edge acoustic modes, where we can reproduce the correct negative sign of the Grüneisen parameter ---which represents a softening of the mode upon compression--- only for InP. Again, this deficiency may be fixed by the introduction of bond charges.

## 8 Results for Nanocrystals

To verify the applicability of the new potentials to nanostructures we calculated the vibrational properties of $In_{225}As_{240}$, $In_{321}P_{312}$, $Ga_{321}P_{312}$, and $Ga_{225}As_{240}$ QDs using DFT and the new empirical potential. For the vibrational properties, these represent the largest clusters addressed at DFT-level up to now. To obtain vibrational properties using the finite difference (or small displacement) scheme, $3N/S$ self-consistent calculations need to be performed, where $N$ is the number of atoms and $S$ the number of symmetry operations of the point group. It is necessary to use these large (for DFT standards) clusters, since the empirical potential needs to be tested at the size regime where they will be used. These QD sizes, although at the limit of modern DFT, represent the smallest realistic structures. It is not useful, and would indeed be detrimental, to test or tune our empirical potential to small, molecule-like clusters.

The results of the vibron DOS are shown in Fig. 4 as black thick lines for the empirical results and as grey (red in color) thin lines for the DFT results. The discrete vibron frequencies are broadened with a Gaussian of width 0.8 cm$^{-1}$ to simplify the comparison. The initial geometry of the QDs (before relaxation) is obtained in a first step by cutting the bulk zinc blende structure into a spherical nanocrystal centered around the cation (Ga or In). In a subsequent step, surface atoms with only one 1NN bond are removed. After this procedure, the nanoclusters retain the $T_d$ point group symmetry derived from bulk zinc blende.

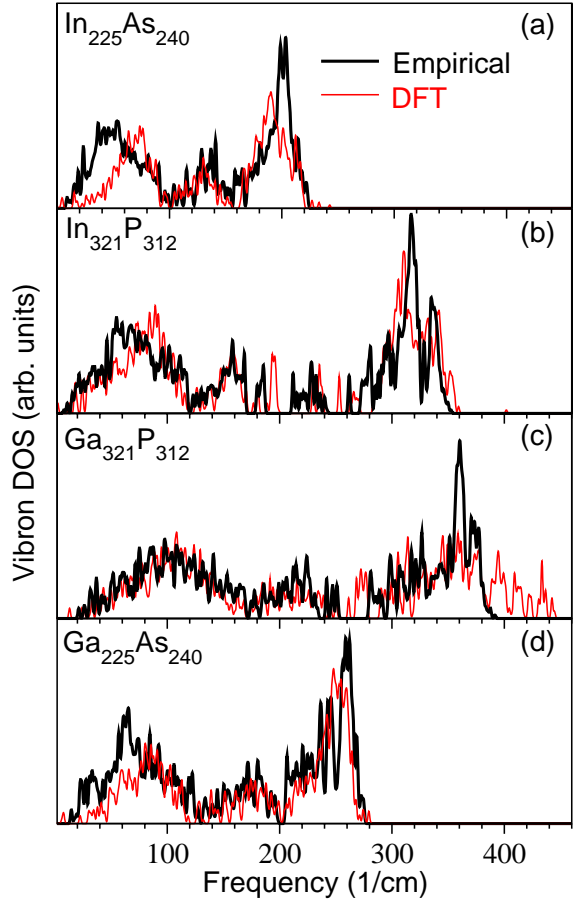

**Figure 4: (Color online) Comparison of the vibron DOS obtained with our empirical potentials (thick lines) and from DFT calculations (thin red lines) for (a) $In_{225}As_{240}$, (b) $In_{321}P_{312}$, (c) $Ga_{321}P_{312}$, and (d) $Ga_{225}As_{240}$ QDs.**

For the empirical relaxation, the surface is unpassivated and no constraints are used on the geometry. The relaxed final structure shows no imaginary frequency, as expected from a properly relaxed structure.

The thin lines in Fig. 4 are the vibron DOS obtained by DFT implemented in CPMD. The DFT calculation is performed within the LDA with norm-conserving pseudopotential at the $\Gamma$ point with an energy cutoff of 30 Ry. The supercell is simple cubic with extent of 36 Å in each direction. In DFT, we perform a symmetry constrained structural relaxation. The unpassivated clusters have a strong tendency to reconstruct[48] and forcing them into an unreconstructed final structure leads to a slow convergence of the relaxation procedure and to a metastable final ``relaxed'' structure. This naturally leads to a large number of imaginary frequencies. However, the tendency to reconstruct in our rather large III-V clusters is very low in the case of a passivated surface. Hence, we relax the structure of the passivated QDs and then remove the passivants to compare the results with the empirical calculations. For the passivants, we use pseudo-hydrogen atoms with a fractional charge of 3/4 and 5/4, passivating the anions and cation, respectively. For the final ``frozen'' geometry we obtain less than 5% imaginary frequencies. The appearance of imaginary frequencies is expected, since the structures are not fully relaxed: after the removal of passivants, forces reappear. In the case of the $Ga_{321}P_{312}$ QD, more than 10% imaginary modes appear. To avoid missing too much information in the vibron spectra, we decided to relax the unpassivated structure, leading to a new structure with less than 2% imaginary modes. Consequences of this procedure are commented below.

We now compare the vibron DOS calculated by DFT and our empirical potential in Fig. 4. First, we analyze the main features describing the effect of confinement on the phonon DOS. We notice 1) The maxima in the bulk DOS corresponding to the van Hove singularities of the acoustic and optical branches remain identifiable in the case of QDs. 2) Compared to the bulk, these are only slightly shifted in frequency (for the cluster sizes considered here). 3) The transition from the crystals space group ($T_d^2$ or $F\bar{4}3m$) to the cluster's point group ($T_d$) leads to mixing of the vibrons derived from different bulk phonon branches. This leads to a broadened phonon DOS compared to the bulk. Especially the sharp and distinct optical branches in bulk merge and broaden in the cluster. 4) Vibrons appear in frequency regions where no vibrations were allowed in bulk, most prominently between the acoustic and optic branches. These vibrons have dominant surface character.

These main features are reproduced by the empirical potential. However, we notice some shift of the lowest frequency maximum for the InAs, GaP and GaAs clusters. This is the consequence of our lower TA frequency at the X-point in bulk, as already pointed out. The deviation at high frequency in the case of the GaP QD is related to the special procedure we used to avoid too many imaginary frequencies in the DFT calculations. We relaxed the unpassivated structures, which leads to a ``buckled'' surface, where the atoms move inwards and partially reconstruct. A full reconstruction would require allowing the system to transform to a lower symmetry structure, which we do not. However, some symmetry conserving surface dimers [48] form. In this case, the shorter bond-length in the surface layer leads to the appearance of higher frequencies. As we do not attempt to model details of the reconstruction and relaxation of a free unpassivated surface, we do not see this as a problem. The high frequency modes of the other QDs (a,b,d) agree well with our model potential.

The motivation for the derivation of empirical potentials is to be able to address the relevant size range of manufactured QDs. In Figure 5 (a)-(l) we show results for the vibron DOS of QDs with diameters $D =$ 32, 45, and 55 Å. The thin gray (red in color) lines in Fig. 5 show the phonon DOS obtained for the respective bulk material.

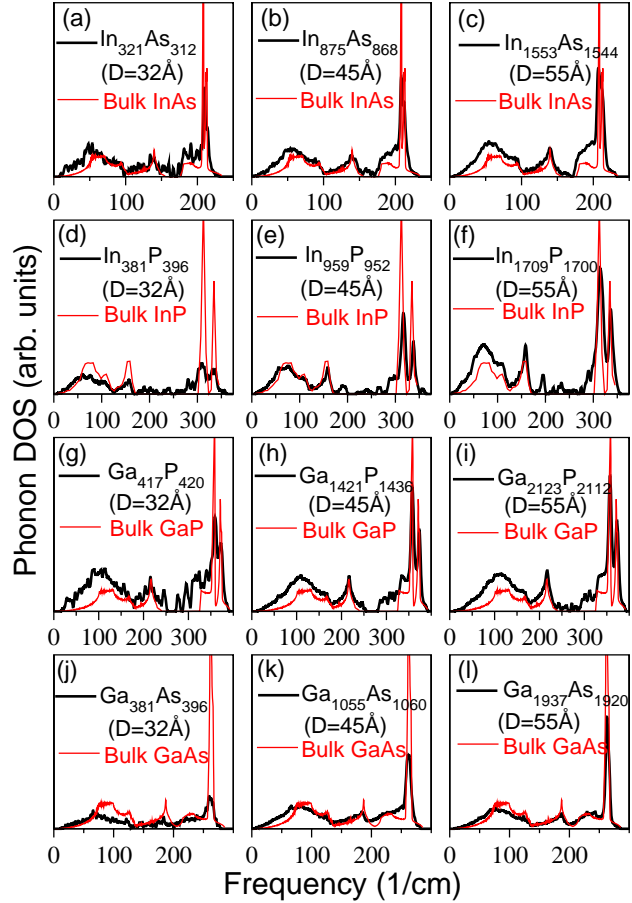

**Figure 5: (Color online) The phonon DOS of QDs (thick lines) and bulk (thin lines) calculated with the modified SW potentials for (a) $In_{321}As_{312}$, (b) $In_{875}As_{868}$, (c) $In_{1553}As_{1544}$, (d) $In_{381}P_{396}$, (e) $In_{959}P_{952}$, (f) $In_{1706}P_{1700}$, (g) $Ga_{417}P_{420}$, (h) $Ga_{1421}P_{1436}$, (i) $Ga_{2123}P_{2112}$, (j) $Ga_{381}As_{396}$, (h) $Ga_{1055}As_{1060}$, and (i) $Ga_{1937}As_{1920}$.**

With increasing QD size the results naturally tend towards the bulk DOS as an increasing fraction of the vibrons have bulk character. Especially the larger GaAs cluster phonon DOS resembles the bulk. This observation may be deceiving as the number of surface related vibrons is still significant, however at this size regime the frequency shifts from the bulk case are shown to be small.

## 9   Summary and Conclusion

In summary, we have derived empirical interatomic potentials for the calculation of the vibrational properties of InAs, InP, GaAs and GaP nanostructures. We introduced a 2NN interaction that effectively screens the long-range Coulomb interaction and allows us to fully relax the structures before we calculate vibrational modes and frequencies. This approach allows to study surface relaxation effects and most importantly leads to solutions free of imaginary frequencies. The potential has the functional form suggested by Stillinger and Weber, which allows the use of most of the empirical interatomic potential codes, without changes. We have generated the potential based on ab-initio DFPT and finite difference calculations for bulk and clusters with up to 633 atoms, respectively. The elastic properties of the bulk material enter the fit as well, which is the key for a reasonable structural relaxations. We find a good agreement between the empirical and ab-initio results, including the broadening of the optical branches and the appearance of surface vibrons in bulk-forbidden frequency ranges. We further show that the vibron DOS tends towards the bulk phonon DOS for clusters with 55Å diameter.

We would like to thank Thomas Hammerschmidt for helpful discussions and acknowledge financial support by the Marie Curie Reintegration Grant. Most of the simulations were preformed on the national supercomputer NEC Nehalem Cluster at the High Performance Computing Center Stuttgart (HLRS).